\documentclass[reprint,aip,pof,superscriptaddress,amssymb, amsmath]{revtex4-1}
\usepackage{mathrsfs}
\usepackage{natbib}

\usepackage{graphicx}
\usepackage{amsfonts}
\usepackage{color}
\usepackage{bm}
\usepackage{import}
\usepackage[english]{babel}

\usepackage{siunitx}

\newcommand{\derd}[2]{\frac{\mathrm{d} #1}{\mathrm{d} #2}}
\newcommand{\derp}[2]{\frac{\partial #1}{\partial #2}}
\renewcommand{\vec}[1]{\bm{#1}}
\renewcommand{\tensor}[1]{\mathbb{#1}}

\newcommand{\ex}{\vec{e}_\mathrm{x}}
\newcommand{\ey}{\vec{e}_\mathrm{y}}
\newcommand{\ez}{\vec{e}_\mathrm{z}}
\newcommand{\ee}{\vec{e}}
\newcommand{\vrr}{\vec{r}}
\newcommand{\va}{\vec{a}}
\newcommand{\vb}{\vec{b}}
\newcommand{\vc}{\vec{c}}

\newcommand{\ve}{\vec{e}}

\newcommand{\ct}{\cos \orient}
\newcommand{\st}{\sin \orient}
\newcommand{\cct}{\cos^2 \!\orient}
\newcommand{\sst}{\sin^2 \!\orient}
\newcommand{\impos}{\vec{x_0}^\star}
\newcommand{\xsing}{\vec{x}_0}
\newcommand{\visc}{\eta}
\newcommand{\orient}{\theta}
\newcommand{\vrat}{\lambda}
\newcommand{\xx}{x}
\newcommand{\yy}{y}
\newcommand{\zz}{z}
\newcommand{\nab}{\nabla_0}

\newcommand{\gs}{\vec{G_S}}
\newcommand{\gsd}{\vec{G_{SD}}}
\newcommand{\gss}{\vec{G_{SS}}}
\newcommand{\gssd}{\vec{G_{SSD}}}
\newcommand{\gsq}{\vec{G_{SQ}}}
\newcommand{\gr}{\vec{G_R}}
\newcommand{\grd}{\vec{G_{RD}}}
\newcommand{\gpd}{\vec{G_{D}}}
\newcommand{\gpq}{\vec{G_{Q}}}

\newcommand{\Eg}{\frac{\gamma^2-1}{\gamma^2+1}}
\newcommand{\EE}{\tensor{E}}
\newcommand{\plan}{\parallel}

\newcommand{\four}[1]{\mathcal{F}\left[#1\right]}

\newcommand{\ti}[1]{\tilde{#1}}

\newcommand{\behat}{\beta}
\newcommand{\lgliss}{\ell}

\newcommand{\dz}{1+\lgliss k}
\newcommand{\ddz}{1+2\lgliss k}

\newcommand{\uind}{\vec{u}^\text{ind}}
\newcommand{\omind}{\vec{\Omega}^\text{ind}}
\renewcommand{\r}{R}
\newcommand{\z}{Z}

\newcommand{\vvec}{\vec{u}^{(2)}}
\newcommand{\vvecf}{\ti{\vec{u}}^{(2)}}

\newcommand{\vscalf}{\ti{u}^{(2)}}

\begin{document}
\title{Dynamics of swimming bacteria at complex interfaces}

\author{Diego Lopez}
\email{diego.lopez@univ-amu.fr}
\affiliation{Aix Marseille Universit\'e, CNRS, IUSTI UMR 7343, 13013 Marseille, France}
\author{Eric Lauga}
\email{e.lauga@damtp.cam.ac.uk}
\affiliation{Department of Applied Mathematics and Theoretical Physics, Centre for Mathematical Sciences, Wilberforce Road, Cambridge, CB3 0WA,United Kingdom}%
\date{\today}
%\revised{}

\begin{abstract}

Flagellated bacteria exploiting helical propulsion are known to swim along circular trajectories near surfaces. Fluid dynamics predicts this circular motion  to be clockwise (CW)  above a rigid surface (when viewed from inside the fluid) and counter-clockwise  (CCW) below a free surface. Recent experimental investigations showed that complex physicochemical processes at the nearby surface could lead to a change in the direction of rotation, both at  solid surfaces absorbing   slip-inducing polymers and interfaces covered with surfactants. Motivated by these results,  we use a far-field hydrodynamic model to predict the kinematics of swimming near three types of interfaces: clean fluid-fluid interface, slipping rigid wall, and a fluid interface covered by incompressible surfactants.  Representing the helical swimmer by a superposition of hydrodynamic singularities, we first show that in all cases the surfaces reorient the swimmer parallel to the surface and attract it, both of which are a consequence of the Stokes dipole component of the swimmer flow field. We then show that circular motion is induced by a higher-order singularity, namely a rotlet dipole, and that its rotation direction (CW vs.~CCW) is  strongly affected by the boundary conditions at the interface and the bacteria shape. Our results  suggest thus that the hydrodynamics of complex interfaces provide a mechanism to selectively  stir bacteria.

\end{abstract}

\maketitle

\section{Introduction}

Swimming microorganisms are ubiquitous in nature, and have long been known to play important roles in marine life ecosystems, animal reproduction, and infectious diseases. In these processes, cell motility is crucial.\cite{braybook} At the small scales relevant to swimming cells, inertial forces are  negligible, and locomotion is constrained by Purcell's ``scallop'' theorem stating that any body deformation reversible in time yields zero net motion.\cite{purcell_1977} Fluid-based cellular motility relies therefore  on non-time reversible deformation, for instance by propagating waves along cilia or flagella.\cite{lauga_2009} 

Among the  various types of locomotion seen in nature, one commonly observed for bacteria  is that of helical propulsion, where a flagellum (or a bundle of flagella) rotates as a helix, inducing forward propulsion. 
 A typical example of an organism employing helical propulsion is  the bacterium \textit{Escherichia coli} ({\it E. coli}).\cite{berg_2004} This bacterium alternates ``run'' and ``tumble'' periods: in the former, flagella are synchronized in a coherent bundle and propel the cell forward, whereas in the latter flagella are disorganized,  changing the cell orientation and subsequent swimming direction. 
During  run periods, when {\it E. coli} cells are
 isolated in a bulk flow, they swim in straight (noisy) lines. 
 
However, cell  locomotion is strongly affected by nearby  boundaries. 
Swimming microorganisms often evolve in confined environments, be it by solid boundaries, free surfaces, or liquid interfaces. In some cases, confinement results from channel boundaries, for example along the mammalian female reproductive tract.\cite{fauci06} Surfaces can also be a key element in the microorganism function, as in the case of surface associated infection or biofilm formation.\cite{lauga_2009,costerton_1995} 
Since such problems are dominated by viscous dissipation, long-range hydrodynamic interactions have been argued to play important roles, resulting in a significant alteration of the locomotion of microorganisms.\cite{lauga_2009} Over the past years, intensive theoretical, numerical and experimental work has helped uncover the  kinematics and dynamics modifications of swimming properties by boundaries.\cite{fauci_1995,harshey_2003, drescher_2011, shum_2010, denissenko_2012}

\begin{figure}[tb]
	\centering
		\includegraphics{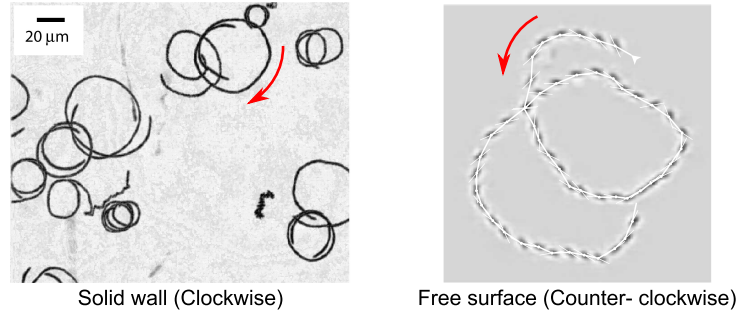}  
	\caption{Experimental evidence of clockwise  motion for bacteria near a solid wall (left panel) [reprinted with permission from E. Lauga, W. R. DiLuzio, G. M. Whitesides, and H. A. Stone, Biophys. J. 90, 400-412 (2006). Copyright (2006) by the Biophysical Society], and counter-clockwise motion at a free surface (right panel) [reprinted figure with permission from R. Di Leonardo, D. Dell'Arciprete, L. Angelani, and V. Iebba, Phys. Rev. Lett. 106, 038101 (2011). Copyright (2011) by the American Physical Society].}
	\label{fig:1}
\end{figure}

For bacteria employing helical propulsion (such as {\it E. coli}), two different effects induced by boundaries have been discovered and quantified. 
These organisms swim in the forward direction (cell body forward) and are being propelled from the back. They thus  push on the surrounding fluid forward and backward, and  such swimmers are referred to as ``pushers''.
In the presence of a nearby solid wall, {\it E. coli} tends to aggregate close to walls.\cite{rotshild_1963} 
This is in fact observed for any kind of pusher, not necessarily one exploiting  helical propulsion.\cite{berke_2008,smith_2009,giacche_2010,maxey_2011} A second property, observed solely for  helical swimmers, is a circular motion of the cells in a plane  parallel to the surface. This was accounted for both experimentally and theoretically in the case of a solid wall  \cite{berg_1990,lauga_2006} and a free surface.\cite{lemelle_2010,dileonardo_2011} Notably, the circular motion occurs in an opposite direction in the presence of a solid wall (clockwise,  CW,  when viewed from inside the fluid) or a free surface (counterclockwise, CCW, see  Fig.~\ref{fig:1}). This change in rotation direction is qualitatively similar to the drag increase or decrease observed for the motion of a colloidal particle  near a rigid  wall and  a free surface.\cite{kim_1991}  Indeed, a solid wall and a free surface induce opposite effects, no-slip for a rigid boundary vs.~free slip in the case of a free interface.

%[width = 0.85\textwidth]

Past experimental  results have been explained theoretically considering Newtonian fluids and perfect interfaces, meaning either  a no-slip  wall  or a shear-free surface. Theoretical models do predict a single circular direction,  CW in the presence of a solid wall vs.~CCW in the presence of a free surface, and are consistent with   the results illustrated in Fig.~\ref{fig:1}. 
However, recent experiments on {\it E. coli} swimming near  glass plates and free surfaces show that the distinction in the direction of the circular motion is not straightforward, and both CW and CCW rotations are observed under seemingly similar experimental conditions.\cite{lemelle_2010,lemelle_2013,morse_2013} In the initial study of Lemelle {\it et al.} (2010),\cite{lemelle_2010}  only CW motion was  observed above a glass plate, but both CW and CCW at a free surface, suggesting that particles and surfactants could alter the free slip boundary condition. This hypothesis was further investigated by changing the concentration of a particular polymer that can aggregate at a free surface.\cite{morse_2013} The authors confirmed this qualitative change of behavior, observing a clear dependence on the polymer concentration of the fraction of cells undergoing CCW motion.
A similar change in rotation direction was recently highlighted experimentally at a solid wall, when the solution contains polymers.\cite{lemelle_2013} Using a special surface treatment, the polymer concentration at the solid wall was modified, generating possible slip, and resulting in CCW motion.
These recent experiments demonstrate that  the presence of polymers or surfactants could have a dramatic effect on  motility of nearby cells. In this paper we present a modeling approach to quantify the dynamics of swimming bacteria near complex interfaces.

When polymers are present in the solution, their concentration close to surfaces is reduced due to higher shear and confinement.\cite{barnes_1995,sochi_2011} This wall depletion results in the formation of a thin fluid layer of lower viscosity at the wall, thereby modifying significantly the no-slip condition. On scales larger than this thin layer, the equivalent behavior at the wall is an apparent partial slip, characterized by its slip length $\lgliss$ ranging from   $\lgliss \sim 10$~nm to  10~$\mu$m.\cite{barnes_1995,mhetar_1998,servantie_2008,lemelle_2013} 
Similarly, a liquid interface covered with surfactants acts as a thin two-dimensional fluid layer separating the liquid phases. This layer has its own rheological properties, and modifies the stress and velocity jumps between the two fluids.\cite{bos_2001,fuller_2003,fuller_2012} As a consequence, the presence of surfactants can affect significantly the boundary conditions and resulting flow.\cite{loewenberg_1999,wang_2013}

In the present work, we address the role of altered  boundary conditions on swimming microorganisms, focusing  on interface-induced reorientation, attraction vs.~repulsion by the surface, and the impact on circular motion. 
Using an analytical framework based on multipole expansions for describing the hydrodynamic interactions between a swimming microorganism and an interface, we show how complex interfaces affect hydrodynamic interactions, providing possible explanations to past experimental results. Whereas interface alignment and attraction are seen to be universal properties, the direction of the circular motion turns out to strongly depends on the properties of the fluid, on the  bacterium shape and, in some cases, the distance to the interface.

In Sec.~\ref{sec:modeling} we present the modeling approach used throughout the paper. We first introduce the different interfaces considered, and then our  solution method quantifying the leading-order effect of hydrodynamic singularities. In Sec.~\ref{sec:StokesletCloseToABoundary} we recall some existing results for the flow generated by a point force near boundaries and derive in particular the solution in the case  of a surfactant-covered interface.   Sec.~\ref{sec:resultstruetrue} is devoted to the main results of the paper, quantifying the  impact of complex boundary conditions  on swimming bacteria, first on reorientation and attraction, and then on circular motion. We finally conclude in  Sec.~\ref{sec:ccl} while some of the technical details are given in Appendices \ref{sec:appB}-\ref{sec:TypicalDifferentialEquationsAndSolutions}.

\section{Modeling}
\label{sec:modeling}
\subsection{Interfaces and boundary conditions}
\label{sec:interfaces}

\begin{figure}[tb]
	\centering
		\includegraphics{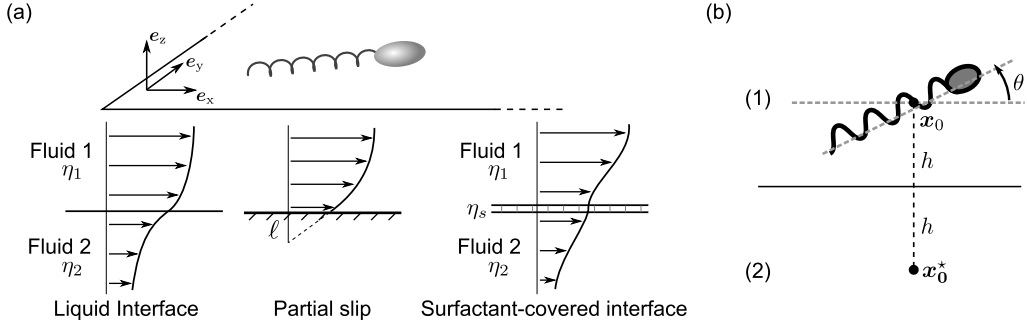}
	\caption{Schematic representation of a swimming bacterium  close to a liquid interface: (a) three different interfaces considered in this work and sketch for their  flow profiles; (b) notations for the calculations in this paper.}
	\label{fig:2}
\end{figure}

Throughout the paper, we use the word interface to  refer to any boundary separating two phases. We denote phase 1 the fluid where the swimming bacterium is located, while  the second phase can either be a solid, a liquid, or a gas.  The presence of a nearby interface affects bacteria locomotion in different  ways depending on its flow  boundary conditions. In this work, we consider three types of interfaces, sketched in Fig.~\ref{fig:2}a: (i) a clean interface, i.e.~a fluid-fluid interface with no surfactant, (ii) a flat surface with a partial slip condition as a model for a polymer depletion layer, and (iii) a liquid interface covered with incompressible surfactants.
In the case of  flow generated by swimming microorganisms, the typical capillary number, scaling the flow-induced stress with  surface tension, is  very small, typically lower than $10^{-5}$. As a consequence we will neglect any  interfacial deformation  induced by the microorganisms, and focus  therefore on planar interfaces with normal   $\ez$ (Fig.~\ref{fig:2}a).

%[width = 0.95\textwidth]

\subsubsection{Clean interface}
The first type of interface we will consider is a clean  interface between two fluids. This problem is characterized by the viscosity ratio between the two fluids,
\begin{align}
	\vrat = \frac{\visc_2}{\visc_1},
\end{align}
where the reference viscosity $\visc_1$ is the viscosity of the fluid where the bacterium is located.
If the viscosity of the second fluid vanishes, $\vrat =0$, the interface behaves as a free surface, whereas  if the second fluid viscosity tends to infinity, we have $\vrat \to \infty$, and 
 the interface becomes equivalent to a rigid wall. A clean liquid interface imposes impermeability and the continuity of tangential velocities and stresses.\cite{blake_1978} The corresponding boundary conditions written at $z=0$ are
\begin{align}
	&u_\xx^{(1)} = u_\xx^{(2)},\quad u_\yy^{(1)} = u_\yy^{(2)},\quad u_\zz^{(1)} = 0,\quad u_\zz^{(2)} = 0,\\
	& \EE_{\xx,\zz}^{(1)}= \vrat \tensor{E}_{\xx,\zz}^{(2)},\quad \tensor{E}_{\yy,\zz}^{(1)}= \vrat \tensor{E}_{\yy,\zz}^{(2)},
\end{align}
where $\EE = \frac{1}{2}(\nabla \vec{u} + \nabla\vec{u}^T)$ is the symmetric  rate of strain tensor, and where we have used the superscript $(i)$ to denote fluid $(i)$. In the case of a free surface ($\vrat=0$), these equations reduce to a free slip condition, whereas in the limit of a rigid wall ($\vrat\to\infty$), the boundary imposes a no-slip condition, $\vec{u}(z=0)=\vec{0}$.

\subsubsection{Slip surface}

The second type of interfaces that we will consider models a rigid wall when the fluid contains polymers, and therefore is subjected to wall depletion. 
For simplicity, we focus solely on the  modification of the boundary conditions and assume  no  change in the fluid rheology. The model problem will therefore become  that of a bacterium swimming in a Newtonian fluid close to a partial-slip boundary, meaning that the wall velocity is proportional to the normal velocity gradient. Defining $\lgliss$ the slip length, the boundary conditions read
\begin{align}
	\vec{u}_\plan = \lgliss\derp{\vec{u}_\plan}{z},\quad u_z = 0,
\end{align}
where $\vec{u}_\plan = u_\xx\ex+u_\yy\ey$ is the flow velocity in the plane of the interface. 

\subsubsection{Surfactant-covered interface}

The last type of interface we are interested in characterizing  are those covered with surfactants. {This type of interface behaves as an  impermeable,   bi-dimensional fluid layer separating the two phases, and the bulk stress jump in the absence of surface forces is given by
\begin{align}
	\derp{}{t}(c_s\vec{u}_\plan) - \nabla_\plan.\vec{\tau}_s  = \ez.[\vec{\tau}^{(1)}-\vec{\tau}^{(2)}],
\end{align}
where $c_s$ is the surfactant concentration, $\vec{u}_\plan$ the surface velocity, $\nabla_\plan$ the in-plane gradient operator, and $\vec{\tau}$ (resp.~$\vec{\tau}_s$) the bulk (resp.~surface) stress tensor.\cite{fuller_2003} 
In the present analysis, it is appropriate to focus on the  stationary limit on the length scales of  swimming bacteria. 
Additionally, we assume that the concentration of surfactant on the surface is large enough so that the relative change in surface tension is of order 1. In that case, as the capillary number is very small (typically lower than $10^{-5}$), the surfactant concentration can be considered to be uniform on the surface.\cite{loewenberg_1999}
Under this assumption, the interface is incompressible, $\nabla_\plan \cdot\vec{u}_\plan = 0$, and the surface stress tensor reads
\begin{align}
	\vec{\tau}_s = \sigma \tensor{I}_s+2\visc_s \tensor{E}_\plan,
\end{align}
where $\sigma$ is the surface tension and $\tensor{E}_\plan$ the in-plane rate of strain tensor.\cite{stone_1990,loewenberg_1999} The resulting boundary conditions at $z=0$ state continuity of in-plane velocity, no  flow perpendicular to the surface, in-plane incompressibility, and balance between shear stresses from the outside flows, surface viscous stresses and Marangoni stresses 
\begin{align}
	&u_\xx^{(1)} = u_\xx^{(2)},\quad u_\yy^{(1)} = u_\yy^{(2)},\quad u_\zz^{(1)} = 0,\quad u_\zz^{(2)} = 0, \\
	&\derp{u_\xx^{(1)}}{x}+\derp{u_\yy^{(1)}}{y} = 0,\\
	&2 \vrat \tensor{E}_{\xx,\zz}^{(2)} - 2\EE_{\xx,\zz}^{(1)}  - \beta h  \nabla_\plan^2u_\xx^{(1)} - \frac 1{\eta_1}\derp{\sigma}{x}=0,\\
	&2 \vrat \tensor{E}_{\yy,\zz}^{(2)} - 2\EE_{\yy,\zz}^{(1)}  - \beta h  \nabla_\plan^2u_\yy^{(1)} - \frac 1{\eta_1}\derp{\sigma}{y}=0,
\end{align}
where $\beta$ is the non-dimensional surface viscosity,
\begin{align}
	\beta= \frac{\visc_s}{\visc_1h},
\end{align}
$h$ being a flow length scale, chosen here to be the bacterium distance to the wall.\cite{loewenberg_1999} The parameter $\beta$ is often referred to as Boussinesq number, comparing surface to bulk stresses \cite{fuller_2003}.}
Note that the condition for incompressibility prevents  the limit of a clean interface to be  reached by  simply applying  $\beta \to 0$, and only the case of a no-slip boundary can be recovered in the limit $\beta \to \infty$.

\subsection{Impact of interface on swimming dynamics}
In order to analyze the effect of an interface on the swimming microorganism, we decompose the flow into two contributions, one being the flow generated by the same organism in the absence of interface, $\vec{U}$, and a contribution due solely to the interface, $\vec{u}^\star$, so that $\vec{u} = \vec{U} + \vec{u}^\star$.
The effect of the interface on the swimming microorganism is then determined using Fax\'en's law{,  modeling the bacterium shape as that of a prolate ellipsoid swimming along an intrinsic direction $\ee$. Noting $\gamma$ the aspect ratio of the entire organism (i.e.~cell body and flagellar bundle),} the interface-induced velocity, $	\uind$,  and rotation rate, $	\omind$,  of the microorganism are given by
\begin{align}
	\label{eq:faxu}
	\uind& =\vec{u}^\star(\vec{x} = \xsing),\\
	\label{eq:faxw}
	\omind& = \vec{\Omega}^\star(\vec{x} = \xsing) = \frac{1}{2}\nabla\times\vec{u}^\star(\xsing) + \Eg~\ee\!\times\!\left[\tensor{E}^\star(\xsing).\ee\right],
\end{align}
where $\xsing$ is the location of the bacterium  \cite{kim_1991} (Fig.~\ref{fig:2}b). 
The fact that the interface effect is evaluated only at the  location of the swimmer is a key point here -- we  do not seek to describe the entire flow field in the presence of the aforementioned interfaces, but only the flow at a specific location.

\subsection{Flow singularities}
\label{sec:flowSing}

The Reynolds numbers associated with swimming microorganisms are very small.\cite{purcell_1977,lauga_2009} In this limit, the flow generated by a swimming microorganism satisfies the linear Stokes equations,\cite{kim_1991} which allow for a multipole expansion of any flow.\cite{chwang_1975} 
In the present work,  we will represent the flow induced by a swimming microorganism  using flow singularities. General  flow singularities and notation are presented below while in  Sec.~\ref{sec:resultstruetrue} we will focus on the specific  singularity model used for microorganisms using helical propulsion.

Flow  singularities are derived from the Green's function for Stokes flows, $\gs$, which gives the flow at position $\vec{x}$ generated by a point force $f$ located at $\xsing$ and oriented along the direction $\va$ as
\begin{align}
	\vec{u}(\vec{x}) &= \frac{f}{8\pi\visc}\gs(\vec{x}-\xsing;\va),\quad \gs(\vrr;\va) = \frac{\va}{r}+\frac{\va.\vrr}{r^3}\vrr,
\end{align}
where $\vrr = \vec{x}-\xsing$, and $r = |\vrr|$. This solution is called the Stokeslet, and higher-order singularities are then derived from this fundamental singularity.\cite{chwang_1975} We introduce here the Stokeslet dipole, $\gsd$,  and quadrupole, $\gsq$, 
\begin{align}
	&\gsd(\vrr;\va,\vb) = (\vb.\nab)\gs(\vrr;\va),\\
	&\gsq(\vrr;\va,\vb,\vc) = (\vc.\nab)\gsd(\vrr;\va,\vb),
\end{align}
and the (potential) source dipole, $\gpd$,  and quadrupole, $\gpq$,
\begin{align}
	&\gpd(\vrr;\va) = -\frac{1}{2}\nab^2\gs(\vrr;\va),\\
	&\gpq(\vrr;\va,\vb) = (\vb.\nab)\gpd(\vrr;\va),
\end{align}
where the notation  $\nab$ is used to denote a gradient taken with respect to the   singularity location, $\xsing$. An useful combination of Stokes dipoles is the rotlet,\cite{chwang_1975} which is the anti-symmetric part of a Stokeslet dipole, and models the flow generated by a point torque
\begin{align}
	\gr(\vrr;\vc) &= \frac{1}{2}\big[\gsd(\vrr;\vb,\va)-\gsd(\vrr;\va,\vb)\big] = \frac{\vc\times\vrr}{r^3},
\end{align}
where $\vc = \va\times\vb$. The symmetric part of a Stokeslet dipole is called a stresslet, $\gss(\va,\vb) =\frac{1}{2}[\gsd(\va,\vb)+\gsd(\vb,\va)]$.
Note that all these singularities are $n$-linear functions of their $n$ orientation vectors. A singularity oriented along arbitrary directions can thus be expressed as a combination of similar singularities along the different basis vectors.

\subsection{Solution method}
\label{sec:solmeth}
The mathematical method used for solving the problem of a point singularity located close to a boundary with specific conditions is Blake's method, presented first for the problem of a Stokeslet close to a rigid wall \cite{blake_1971}.  The effect of the interface  is mathematically  equivalent to an additional flow generated by a system of hydrodynamic  image singularities  located on the other side of the interface. For complex boundary conditions, the image system is a spatial distribution of singularities.

The problem can be significantly simplified by  guessing part of the image system. 
Noting $\text{Im}\{\vec{U}\}$ the image system, the flow is decomposed as 
\begin{align}
	&\vec{u}^{(1)} = \vec{U}+\text{Im}\{\vec{U}\}, \text{ with } \text{Im}\{\vec{U}\}=\vec{V} +\vec{w},\\
	&\vec{u}^{(2)} = \text{Im}^{(2)}\{\vec{U}\},
\end{align}
where $\vec{U}$ is the bulk solution due to the singularity itself, $\vec{V}$ is a guess in the image system, and $\vec{w}$ and $\vvec$ are the unknowns. For example, in the case of a Stokeslet close to a free surface, a good guess for $\vec{V}$ is a Stokeslet symmetric with respect to the interface, which  is enough to enforce both the impermeability condition and that of zero shear stress, resulting in $\vec{w}=\vec{0}$. The major advantage of this decomposition is that the forcing term due to the singularity in the flow equation is solved by the bulk flow term $\vec{U}$. As a result, $\vec{w}$ and $\vvec$ satisfy Stokes equations in the absence of forcing terms
\begin{align}
		\nabla.\vec{u} = 0,\quad \visc \nabla^2 \vec{u} = \nabla p,
		\label{eq:stokes}
\end{align}
where $\vec{u}$ is the velocity field and $p$ the pressure, $\visc$ being the fluid viscosity.
That problem can be more easily solved in Fourier space,\cite{blake_1971,lauga_2005} using a two-dimensional (2D) Fourier transform, defined as
\begin{align}
	\ti{f}(k_1,k_2,z) = \four{f} = \frac{1}{2\pi}\int\int{f(x,y,z)e^{ik_1x+ik_2y}\mathrm{d}x\mathrm{d}y}.
\end{align}
Some useful Fourier transforms are referenced in Appendix \ref{sec:appB}.
The solution of Stokes equations in a 2D Fourier space is straightforward, and reads for $\ti{\vec{w}}$ and $\vvecf$
\begin{align}
	\ti{\vec{w}} = \left(\frac{1}{8\pi\visc_1}\right)	
	\begin{pmatrix}
	A_\xx + ik_1 Bz\\
	A_\yy + ik_1 Bz\\
	A_z + k Bz
	\end{pmatrix}
	e^{-kz}, \quad \vvecf = \left(\frac{1}{8\pi\visc_2}\right)	
	\begin{pmatrix}
	C_\xx + ik_1 Bz\\
	C_\yy + ik_1 Bz\\
	C_z - k Bz
	\end{pmatrix}
	e^{kz}.  
	\label{eq:ufour}
\end{align}
The unknown coefficients in this solution are determined using the continuity equation\begin{align}
	i(k_1A_\xx + k_2A_\yy) = k(B-A_z),  \quad 	i(k_1C_\xx + k_2C_\yy) = k(C_z-D),
\end{align}
together with the relevant boundary conditions. 

Once the solution is obtained for the Stokeslet, it is then possible to derive the solution for higher-order singularities from the Stokeslet solution. This is achieved by applying the same operator acting on the singularity position, $(\vb.\nab)$, where $\vb$ is the direction where the derivative is taken. Note that this can be done more easily in Fourier space, as
\begin{align}
	\four{\ex.\nab}=ik_1,\quad \four{\ey.\nab}=ik_2, \quad \four{\ez.\nab}=\derp{}{h}.
\end{align}
However, it is also possible to directly apply Blake's method to any singularity, which  can be more convenient if a good guess of $\vec{V}$ is found.

\section{Stokeslet close to a complex interface}
\label{sec:StokesletCloseToABoundary}

As a singularity is a linear function of its orientation vectors, one only needs to solve the problem for  the  Stokeslet along the directions parallel and perpendicular to the interface. The flow generated by higher-order singularities can then be derived from these projections.  By  symmetry reasons, all orientations in the plane parallel to the interface are equivalent, we therefore choose $(\ex,\ez)$ to be the singularity plane.
We review below some solutions of the flow generated by parallel and perpendicular Stokeslets close to a fluid boundary, and then use the method described in the previous section for determining the flow in the presence of a surface covered with incompressible surfactant, providing an alternative to the derivation by {B\l awzdziewicz} {\it et al.} (1999).\cite{loewenberg_1999} The solution in the case of slip is given in Lauga and Squires (2005)\cite{lauga_2005} along the same lines.

\subsection{Stokeslet near a clean fluid-fluid interface}
\label{sec:existingSolutions}

The case of parallel and perpendicular Stokeslets were derived by Blake in the case of a solid wall, a free surface, and a fluid-fluid interface.\cite{blake_1971,blake_1978}
In the presence of a clean fluid-fluid interface, the image system is made of a finite number of point singularities located at the image position $\impos$ in fluid 2 (see notation in Fig.~\ref{fig:2}b). The image system is composed of a Stokeslet, a Stokes dipole, and a source dipole, whose intensities vary with the viscosity ratio $\vrat = \visc_2/\visc_1$ and the distance $h$ to the interface. For Stokeslets  parallel, $\gs(\ex)$,    and  perpendicular to the surface, $\gs(\ez)$,    the image system is given by
\begin{align}
	\label{eq:imgsp}
	&\text{Im}\{\gs(\ex)\}= \frac{1-\vrat}{1+\vrat} \gs^\star(\ex)+\frac{2\vrat h}{\vrat+1}\gsd^\star(\ez,\ex)-\frac{2\vrat h^2}{\vrat+1}\gpd^\star(\ex),\\
	\label{eq:imgsper}
	&\text{Im}\{\gs(\ez)\}= -\gs^\star(\ez)-\frac{2\vrat h}{\vrat+1} \gsd^\star(\ez,\ez)+\frac{2\vrat h^2}{\vrat+1}\gpd^\star(\ez),
\end{align}
where we use $\gs(\ve) = \gs(\vrr;\ve)$ in order to simplify the notations, and where the superscript  $^\star$ implies that the singularity is located at the image position, $\impos$. 
These singularities give the flow in fluid 1, where the singularity is located. The flow in fluid 2 is given by a second set of singularities,  given by 
\begin{align}
	&\text{Im}^{(2)}\{\gs(\ex)\}= \frac{2}{1+\vrat} \big[\gs(\ex)+h\gsd(\ez,\ex)+h^2\gpd(\ex)\big],\\
	&\text{Im}^{(2)}\{\gs(\ez)\}= \frac{2 h}{1+\vrat} \big[\gsd(\ez,\ez)+h\gpd(\ez)\big].
\end{align}
This second set of images is located at the singularity position $\xsing$, in fluid 1 (see Fig.~\ref{fig:2}b).
The solutions are therefore obtained fully analytically, and higher-order solutions can be obtained by taking derivatives with respect to the singularity position ($\va.\nab$).
Note that since the image strengths are functions of the singularity distance to the interface $h$, taking derivatives of the image system along $\ez$ generates additional singularities.\cite{spagnolie_2012}

In the presence of a solid wall, the second set of images vanishes since there is no flow in the solid region. This was the no-slip solution originally presented by Blake.\cite{blake_1971}
As we are interested in this work only on the effect of the interface on the swimming bacterium, we will focus only on the first set of images, Eqs.~\eqref{eq:imgsp} and \eqref{eq:imgsper}, giving the flow field in the region where the swimming microorganism is located.

The no-slip and no-shear solutions have been used in the past to explain different swimming behaviors of bacteria  observed  experimentally, such as wall attraction, alignment, and circular motion.\cite{lauga_2006, berke_2008, dileonardo_2011} The particular case of no-slip boundary ($\vrat = \infty$) is noteworthy, and will be used in the following as a reference solution
\begin{align}
	\label{eq:wsp}
	\text{Im}\{\gs(\ex)\} _{\vrat\to\infty} &=  - \gs^\star(\ex)+2h\gsd^\star(\ez,\ex)-2 h^2\gpd^\star(\ex),\\
	\label{eq:wsper}
	\text{Im}\{\gs(\ez)\} _{\vrat\to\infty} &=-\gs^\star(\ez)-2h \gsd^\star(\ez,\ez)+2h^2\gpd^\star(\ez).
\end{align}

In the presence of slip or surfactants, the image system is no longer a set of point singularities. The solution to the problem of a Stokeslet close to a partial slip boundary was addressed by Lauga and Squires,\cite{lauga_2005} using Blake's method. The case of a surfactant-covered interface has been addressed by {B\l awzdziewicz} {\it et al.},\cite{loewenberg_1999} using a Batchelor-type decomposition of the flow. In order to use the same formalism throughout our paper, we derive below the solution of a Stokeslet close to  a surface covered with incompressible surfactants using Blake's method.

\subsection{Stokeslet close to an interface covered with incompressible surfactant}
\label{sec:surfactstokeslet}

We consider here the case of a Stokeslet perpendicular and  parallel  to a surfactant-covered interface as defined in Sec.~\ref{sec:interfaces}. Without loss of generality,  the Stokeslet strength is taken to be 1. 
For this problem, we take  $\vec{V}$ to be the  opposite Stokeslet located at the image position $\impos$. This choice yields at $z=0$ for a perpendicular and parallel Stokeslet respectively
\begin{align}
	\left[\vec{U}+\vec{V}\right](\ez) = -\frac{h}{4\pi\visc_1}\frac{1}{r_h^3}(x\ex+y\ey),\quad \left[\vec{U}+\vec{V}\right] (\ex)= -\frac{h}{4\pi\visc_1}\frac{x}{r_h^3}\ez,
\end{align}
where $r_h^2 = x^2+y^2+h^2$. The problem is then solved in Fourier space.

\subsubsection{Solution for a perpendicular Stokeslet}

In the case of a perpendicular Stokeslet, the boundary conditions read in Fourier space
\begin{align}
	&\ti{w}_\alpha - \vscalf_\alpha  = \frac{h}{4\pi\visc_1}\frac{ik_\alpha}{k}e^{-kh}, \quad \ti{w}_z = 0, \quad \vscalf_z = 0,\\
	&-i(k_1\ti{w}_1 + k_2\ti{w}_2) = \frac{kh}{4\pi\visc_1}e^{-kh},\\
	&\visc_2\left(\derp{\vscalf_\alpha}{z} - ik_\alpha\vscalf_z\right) - \visc_1\left(\derp{\ti{w}_\alpha}{z} - ik_\alpha \ti{w}_z\right) + k^2\visc_s \ti{w}_\alpha +ik_\alpha \ti{\sigma} = \frac{h\visc_s}{4\pi\visc_1} ik_\alpha k e^{-kh},
\end{align}
for $\alpha=1$ and 2, where index 1 (resp.~2) is associated with the $x$ (resp.~$y$) coordinate.
This set of equations is satisfied by $\vvecf = \vec{0}$, which corresponds to a solid wall. We recover therefore a known result:  because of  surface incompressibility,  the flow generated by a Stokeslet perpendicular  to an interface covered with incompressible surfactants is identical to that in the presence of a solid wall.\cite{loewenberg_1999}

\subsubsection{Solution for a parallel Stokeslet}

In this case, the boundary conditions are given  in Fourier space by
\begin{align}
	&\ti{w}_\alpha = \vscalf_\alpha , \quad \ti{w}_z = \frac{h}{4\pi\visc_1}\frac{ik_1}{k}e^{-kh}, \quad \vscalf_z = 0,\quad -i(k_1\ti{w}_1 + k_2\ti{w}_2) = 0,\\
	&\visc_2\left(\derp{\vscalf_1}{z} - ik_1\vscalf_z\right) - \visc_1\left(\derp{\ti{w}_1}{z} - ik_1 \ti{w}_z\right) + k^2\visc_s \ti{w}_1 +ik_1 \ti{\sigma} = 
		\frac{1}{2\pi} \left(1-\frac{k_1^2}{k}h\right)e^{-kh} ,\\
	&\visc_2\left(\derp{\vscalf_2}{z} - ik_2\vscalf_z\right) - \visc_1\left(\derp{\ti{w}_2}{z} - ik_2 \ti{w}_z\right) + k^2\visc_s \ti{w}_2 +ik_2 \ti{\sigma} = 	
		-\frac{h}{2\pi}\frac{k_1k_2}{k}e^{-kh}.
\end{align}
Introducing the coefficients $A$, $B$, $C$ and $D$ from Sec.~\ref{sec:solmeth}, this system can be solved directly, leading to $\ti{\sigma} = ik_1(kh-1)e^{-kh}/(2\pi k^2)$, and
\begin{align}
	&A_\xx  = \frac{4}{1+\lambda+\beta hk}~\frac{1}{k^2}~\left(\frac{k_2^2}{k}e^{-kh}\right),\quad A_\yy  =\frac{4}{1+\lambda+\beta hk}~\frac{1}{k^2}~\left(\frac{-k_1k_2}{k}e^{-kh}\right), \\
	&C_\xx  = A_\xx,\quad C_\yy  = A_\yy, \quad A_z  = B = 2h\frac{ik_1}{k}e^{-kh},\quad	C_z =D = 0.
\end{align}
It is interesting to note that the coefficients in the no-slip problem read
\begin{align}
	A^0_\xx = 0, \quad A^0_\yy = 0, \quad A^0_\zz = 2h\frac{ik_1}{k}e^{-kh}, \quad B^0 = 2h\frac{ik_1}{k}e^{-kh}.
\end{align}
Hence, introducing $\ti{\vec{W}}(k_1,k_2,z) = [A_\xx(k_1,k_2) \ex + A_\yy(k_1,k_2) \ey]e^{-kz}/(8\pi\visc_1)$, the flow field $\vec{w}$ can be written as
\begin{align}
	\vec{w} = \vec{w}^0 + \vec{W},
\end{align}
where $\vec{w}^0$ is the known solution of the problem in the presence of a no-slip boundary, $\vec{w}^0 = 2h\gsd^\star(\ez,\ex)-2h^2\gpd^\star(\ex)$.
After an inverse Fourier transform of $\ti{\vec{W}}$, one can identify two differential equations satisfied by $W_\xx$ and $W_\yy$ in real space, which can be written as a single vectorial differential equation on $\vec{W}$, 
\begin{align}
	\left(1+\lambda -\beta h\derp{}{z}\right)\derp{^2\vec{W}}{z^2}= 4\grd^\star(\ez,\ey),
	\label{eq:wstokeslet}
\end{align} 
where $\grd^\star(\ez,\ey)$ is a dipole along $\ey$ of rotlets along $\ez$, located at $\impos$. 
The  rotlet dipole involved in this flow corresponds to two vertical counter-rotating point vortices, merging at $\impos$.
Note that this singularity can also be written in a way similar to that of Ref.\cite{loewenberg_1999} as $\grd^\star(\ez,\ey) = \ez\wedge \gpd^\star(\ey)$.
The flow in the second fluid satisfies a similar problem,
\begin{align}
	\left(1+\lambda -\beta h\derp{}{z}\right)\derp{^2\vvec}{z^2} = 4\grd(\ez,\ey),
\end{align}
with the rotlet dipole being located at $\xsing$.
As a result, the total flow generated by a Stokeslet parallel to an interface covered with surfactants can be written as
\begin{align}\label{avec0}
	\vec{u}_S(\ex)  = \vec{u}^0_S(\ex) + \vec{W},
\end{align}
where $\vec{u}^0_S(\ex)$ is the flow generated by a Stokeslet along $\ex$ in the presence of a no-slip boundary. We recover therefore the result of Ref.\cite{loewenberg_1999} where the flow is the sum of the no-slip contribution and a ``surface-solenoidal'' flow, i.e.~a 2D flow decaying in $z$.

In order to obtain the exact expression for the flow at the position of the singularity, one still needs to integrate a third order differential equation, Eq.~\eqref{eq:wstokeslet}. 
Since our goal is not to derive the entire flow but only the effect of the boundary on the swimming microorganism, we can simplify in the following way. The singularity is located at $x=y=0$ and since  the differential equations giving the flow are equations in $z$, it is possible to set $x=y=0$ before integrating.
As a result, the additional flow induced on the singularity is given by $W^\text{ind}_\xx(\z=2h)\ex+W^\text{ind}_\yy(\z=2h)\ey$, where $\z = z+h$ and
\begin{align}
	\left(1+\lambda -\beta h\derd{}{\z}\right)\derd{^2W^\text{ind}_\xx}{\z^2} &= \frac{1}{2\pi\visc_1}\frac{1}{\z^3},\quad	\left(1+\lambda -\beta h\derd{}{\z}\right)\derd{^2W^\text{ind}_\yy}{\z^2} = 0.
\end{align}
The latter equation yields $W^\text{ind}_\yy=0$, which could have been anticipated by  symmetry. The former equation can be integrated analytically, as detailed in the  next section.

\renewcommand{\uind}{u^\text{ind}_\zz}
\renewcommand{\omind}{\Omega^\text{ind}}

\section{Swimming microorganisms and interfaces}
\label{sec:resultstruetrue}

\subsection{Bacteria modeled as point singularities}

In this work, we model a swimming bacterium in the far field and thus assume that it is  small compared to any flow length scale. In particular, the distance between the cell and  the interface, $h$, must remain larger than the bacterium size. This far-field approach allows for a representation of a swimming microorganism as a combination of point singularities, while  the organism size and shape play a role only through the aspect ratio $\gamma$ involved in Fax\'en's law, Eq.~\eqref{eq:faxw}.

\begin{figure}[tb]
	\centering
		\includegraphics{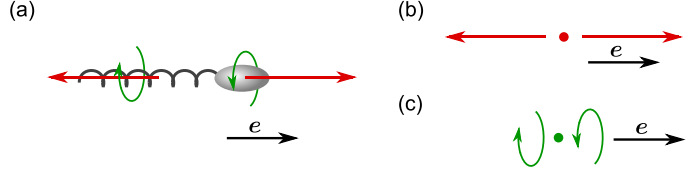}
	\caption{(a) The forces generated by a swimming bacterium such as {\it E. coli} include two flow singularities which impact the wall dynamics: (b) a force dipole due to thrust from the flagella and drag on the body and (c) a rotlet dipole arising from the counter-rotation between body and flagella.}
	\label{fig:3}
\end{figure}
%[width = 0.7\textwidth]

We focus here on microorganisms using helical propulsion, like \textit{E.~coli}, as illustrated schematically in Fig.~\ref{fig:3}a. Such micro-swimmers are force- and torque-free, and can be considered as axisymmetric in average. 
We choose as a convention that the microorganism swims in the $(\ex,\ez)$ plane, along direction $\ee = \ct~\!\ex+\st~\!\ez$.
At leading order (spatial decay $1/r^2$), the far-field flow generated by a swimming bacterium is well captured by an axisymmetric Stokeslet dipole, $\gsd(\ee,\ee)$, as sketched in Fig.~\ref{fig:3}b.\cite{drescher_2011} This singularity is force- and torque-free, and accounts for the spatial distribution of thrust and drag on a flagellated swimming cell.\cite{lauga_2009} This representation is commonly used for describing  far-field hydrodynamic interactions, both  to quantify collective dynamics \cite{simha_2002,hernandez_2005,saintillan_2008,koch_2011} and wall effects.\cite{berke_2008,smith_2009,spagnolie_2012}

However, by symmetry, this leading-order singularity   cannot be responsible for the observed circular rotation along the normal to the interface since $(\ex,\ez)$ is a symmetry plane of the Stokes dipole. It is thus necessary to add higher-order singularities decaying as $1/r^3$ to capture circular swimming. Those singularities could be: (i) a Stokes quadrupole, related to the length asymmetry between the flagella and the body, (ii) a potential source dipole accounting for the finite size of the bacterium, and (iii) a rotlet dipole  capturing the counter-rotation between the flagella and the body. 
By symmetry, it is straightforward to see that the only singularity that could potentially lead to nonzero rotation along $\ez$ is the rotlet dipole $\grd(\ee,\ee)$, illustrated in Fig.~\ref{fig:3}c. 

We are interested in three effects induced by a nearby boundary:  (a) reorientation in the swimming plane, given by rotation rate $\omind_\yy$; (b)  attraction or repulsion by the wall, quantified by wall-induced velocity $\uind$;  and (c)  circular motion, measured by the out-of-plane rotation rate $\omind_\zz$. Reorientation and attraction  can both be accounted for by the leading-order Stokes dipole, as analyzed in Sec.~\ref{lo}. In contrast, the circular swimming which is due to the rotlet dipole is addressed in Sec.~\ref{no}.

\subsection{Leading order: Stokes dipole close to a complex interface}
\label{lo}

\renewcommand{\omind}{\Omega^\text{ind}_\yy}

We  consider here the leading order singularity representing a swimming microorganism, a Stokes dipole, and focus on two effects.  
We first address  the issue of reorientation in the $(\ex,\ez)$ plane  induced by a boundary on a tilted micro-swimmer moving along the direction $\ee = \ct~\!\ex+\st~\!\ez$. With the stable wall-induced orientations derived, $\orient_0$, we then address the attractive vs.~repulsive nature of the interface.

A  Stokes dipole tilted along the angle $\theta$  is a combination of parallel and perpendicular Stokes dipoles as
\begin{align}
	\vec{u}_{SD}(\ee,\ee) &= \cct~\!\vec{u}_{SD}(\ex,\ex) + \sst~\!\vec{u}_{SD}(\ez,\ez) + \sin(2\orient)~\!\vec{u}_{SS}(\ex,\ez).
	\label{eq:tiltdipdecomp}
\end{align}
The wall-induced rotation rate is dependent on the orientation of the microorganism and the local strain rate, see Eq.~\eqref{eq:faxw}.  In the following, the strength of the Stokes dipole is taken to be 1, modeling a ``pusher'' swimmer as relevant to any flagellated bacteria moving cell body first (a puller corresponds to a negative strength).

\subsubsection{Clean interface}
In the presence of a clean interface, the image system is the set of point singularities listed above, and the problem can be solved analytically, leading to the wall-induced rate
\begin{align}
	\omind = \frac{3\ct\st}{64\pi\visc_1h^3}\left[1 +\frac{1}{2}\Eg \frac{\lambda + (2+\lambda)\sst}{1+\lambda}\right].
	\label{omcleansurf}
\end{align}
Notably, the  sign of this rotation rate is given by $\sin(2\orient)$ as the term in the bracket is always positive. As a result, the interface will always tend to align the (pusher) Stokes dipole in the direction parallel to the interface. This result was previously shown in the case of a solid wall,\cite{berke_2008} and is thus generalized here to any fluid-fluid interface.

The stable orientation is therefore a microorganism swimming parallel to the interface, $\orient_0=0$. In that case, the induced velocity along the vertical axis reads
\begin{align}
	\uind =  - \frac{2+3\lambda}{64\pi\visc_1(1+\lambda)h^2}\ez.
\end{align}
We see thus that a swimming pusher will be attracted by a nearby liquid interface for any value of the viscosity ratio. In the limits of a free surface ($\vrat = 0$) and a solid wall ($\vrat=\infty$), standard results are recovered.\cite{berke_2008,spagnolie_2012}

\subsubsection{Partial slip boundary}
\label{sec:partslipdip}

In the presence of a partial slip boundary, the flow due to the boundary is equivalent to that generated by a continuous distribution of singularities along the vertical axis on the other side of the surface.\cite{lauga_2005} Taking derivatives of higher-order solutions is not straightforward in real space, but is  easily carried out in Fourier space.
It is interesting to note that $(\ey,\ez)$  is a symmetry plane for the first two singularities of the decomposition shown in Eq.~\eqref{eq:tiltdipdecomp}, namely $\vec{u}_{SD}(\ex,\ex)$ and $\vec{u}_{SD}(\ez,\ez)$. As a result, these terms do not contribute to any rotation rate through vorticity, only the stresslet does. However, all terms contribute to the total rotation rate, as the swimmer's orientation breaks the symmetry, acting on the strain contribution to the rotation rate, Eq.~\eqref{eq:om_gen}.

The three singularities  necessary for describing a tilted Stokes dipole are derived with a particular choice for the different velocity fields $\vec{V}$, so that $\vec{V}$ corresponds to the solution in the presence of a free surface. This limit is reached in the partial slip model when $\lgliss$ tends to infinity. For the stresslet $\vec{U}_{SS}(\ex,\ez)$, we choose therefore $\vec{V} = -\gss^\star(\ex,\ez)$; the no-slip solution reads 
\begin{align}
	\vec{w}^0 &= 2\gss^\star(\ex,\ez)-2h\gsq^\star(\ez,\ez,\ex) -2h\gpd^\star(\ex)+2h^2\gpq^\star(\ez,\ex).
\end{align}
Following the procedure outlined above  we determine the solution in the presence of a partial slip boundary in Fourier space 
\begin{align}
	\ti{\vec{w}} &=\frac{\ti{\vec{w}}^0}{\ddz} +\frac{\lgliss}{4\pi\mu(\dz)(\ddz)}\left(\frac{k_2^2}{k}\ex - \frac{k_1k_2}{k}\ey\right)e^{-k(z+h)}.
\end{align}
Similarly  we have for the parallel Stokes dipole $\vec{u}_{SD}(\ex,\ex)$ \cite{lauga_2005}
\begin{align}
	\vec{V}=\gsd^\star(\ex,\ex),\quad \ti{\vec{w}} = \frac{\ti{\vec{w}}^0}{\ddz} -\frac{\lgliss ik_1k_2 e^{-k(z+h)}}{2\pi\visc_1(\dz)(\ddz)k^2}\left[k_2\ex-k_1\ey\right],
	\label{eq:sdpar}
\end{align}
with the no-slip solution $\vec{w}^0 = -2\gsd^\star(\ex,\ex) + 2h\gsq^\star(\ez,\ex,\ex) -2h^2\gpq^\star(\ex,\ex)$. 
For the perpendicular Stokes dipole $\vec{u}_{SD}(\ez,\ez)$, we have
\begin{align}
	&\vec{V}=\gsd^\star(\ez,\ez),\quad \ti{\vec{w}} = \frac{\ti{\vec{w}}^0}{\ddz},\\
	&\vec{w}^0 = 2\left[-\gsd^\star(\ez,\ez) + h\gsq^\star(\ez,\ez,\ez) +2h\gpd^\star(\ez)-h^2\gpq^\star(\ez,\ez)\right].
\end{align}

The rotation rate of the cell along $\ey$ is then computed in Fourier space for the three singularities, and the total rotation rate on a tilted Stokes dipole close to a partial slip boundary finally reads
\begin{align}
	\omind = \cct~\!\Omega_{SDX} + \sst~\!\Omega_{SDZ} + \sin(2\orient)\Omega_{SS},
\end{align}
evaluated at the singularity position, where index SDX stands for parallel Stokes dipole, SDZ for perpendicular Stokes dipole, and SS for the stresslet contribution.
The technical difficulty in this problem is the inverse Fourier transform of the rotation rate and we refer to Appendices \ref{sec:appA} and \ref{sec:TypicalDifferentialEquationsAndSolutions} for the details related to the inversion problem and solution of the resulting differential equations. The final solution in real space is given analytically by 
\begin{align}
	\omind & =  \frac{3\ct\st}{64\pi\visc_1h^3}\Bigg\{1+ \frac{2h}{\lgliss}\left(F_4\!\!\left[\frac{h}{\lgliss}\right]-F_3\!\!\left[\frac{h}{\lgliss}\right]-H_4\!\!\left[\frac{h}{\lgliss}\right]\right) +\Eg\sst\nonumber\\
	&\qquad\qquad \quad \quad +\Eg \frac{h}{\lgliss} \bigg[-2\sst F_3\!\!\left[\frac{h}{\lgliss}\right] +(3\cct-2\sst)H_4\!\!\left[\frac{h}{\lgliss}\right]\nonumber\\
	&\qquad \qquad\qquad \qquad \quad+(1+3\sst) \left(F_4\!\!\left[\frac{h}{\lgliss}\right]-\frac 12F_5\!\!\left[\frac{h}{\lgliss}\right]+\frac 12G_4\!\!\left[\frac{h}{\lgliss}\right]-G_5\!\!\left[\frac{h}{\lgliss}\right]\right)\bigg]\Bigg\},
\end{align}
where the functions $F_n$, $G_n$ and $H_n$ derive from the exponential integral function of order $n$, $E_n$, as 
\begin{equation}\label{FnGnHn}
	F_n\!(x) = e^xE_n\!(x),\quad G_n\!(x) =  e^x \int_1^\infty{\frac{E_n\!(x t)}{t^{n-1}}\mathrm{d}t}, \quad H_n\!(x) =  e^{2x} \int_1^\infty{\frac{e^{-x t}}{t^{n-1}}E_n\!(x t)\mathrm{d}t},
\end{equation}
with 
\begin{equation}
\label{En}
	E_n\!(x) = \int_1^\infty{\frac{e^{-x t}}{t^n}\mathrm{d}t}.
\end{equation}

\begin{figure}[tb]
	\centering
		\includegraphics{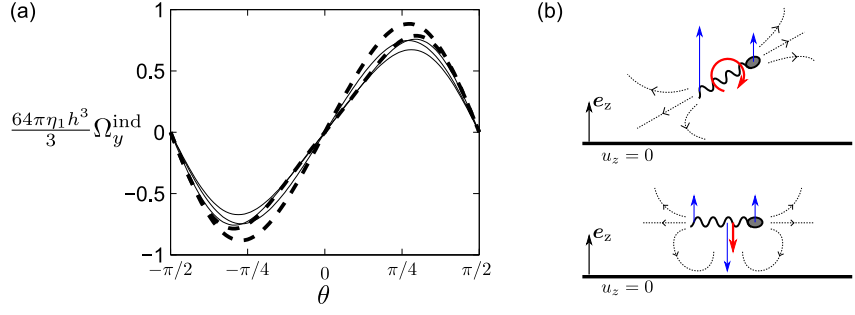}
	\caption{(a) Interface-induced rotation rate along, $\omind$, characterizing the alignment of the  bacterium with the boundary, as a function of the orientation angle $\orient$ and for different values of the dimensionless slip length $\lgliss/h$ (0.1, 1, and 10, thin solid lines). Bold dashed lines correspond to the limits of vanishing slip length (no-slip condition) and infinite slip length (no shear). (b) Qualitative physical picture  for interface alignment and attraction; streamlines are thin dashed lines, thin arrows (dark blue online) show local effects while thick arrows (light red online) show global effect on the swimming microorganism.}
	\label{fig:4}
\end{figure}
%[width=0.8\textwidth]

The dependence of the rotation rate $\omind$ induced by a partial slip boundary on a tilted Stokeslet dipole is shown in Fig.~\ref{fig:4}a, as a function of the tilt angle $\orient$, for different values of the slip length. We see that the slip length, similarly to  the viscosity ratio in the case of a clean liquid interface, does not modify qualitatively the reorientation dynamics and alignment rate with the boundary and a pusher micro-swimmer will always tend to align parallel to the boundary. 
The limit of infinite slip length, $\vrat\to 0$ in Eq.~\eqref{omcleansurf}, leads naturally to the free surface limit. Furthermore, using the asymptotic properties of the exponential integral functions, we have $F_n\!(x)\sim 1/x$ and $G_n\!(x)\sim H_n\!(x) \sim 1/x^2$ when $x$ tends to infinity, which can then be used to recover quantitatively the reorientation dynamics in the no-slip limit.

We then compute  the velocity induced on a micro-swimmer parallel to the boundary, modeled by a parallel Stokes dipole, Eq.~\eqref{eq:sdpar}. The integration can be done in a similar way as for the rotation rate, and using integration by parts we get a simple expression for the vertical velocity as
\begin{align}
	\uind = -\frac 1{32\pi\visc_1h^2}\Bigg\{1 + \frac{h}{4\lgliss}\left( 1 + \frac{h}{\lgliss} + \frac{h}{\lgliss}\left(2+\frac{h}{\lgliss}\right) F_1\!(h/\lgliss) \right)\Bigg\}.
\end{align}
A partial slip boundary will always attract a  pusher swimmer for any value of the slip length, generalizing therefore the known results for $\ell=0$ (no-slip) and $\ell=\infty$ (no-shear). 

\subsubsection{Surfactant-covered interface}

We finally consider  the case of a Stokes dipole close to a liquid interface covered with incompressible surfactants. From the Stokeslet solution presented in Sec.~\ref{sec:surfactstokeslet}, it is possible to derive higher-order solutions. Using Eq.~\eqref{avec0}, the flow generated  by a tilted Stokeslet dipole is therefore given by
\begin{align}
	\vec{u}_{SD}(\orient,\orient) = \vec{u}^0_{SD}(\orient,\orient)+ \cct \vec{W}_{SD}(\ex,\ex) + \ct\st\vec{W}_{SD}(\ex,\ez),
\end{align}
with
\begin{align}
	\vec{W}_{SD}(\ex,\ex) = -\derp{\vec{W}}{x_1}, \quad \vec{W}_{SD}(\ex,\ez) = \derp{\vec{W}}{h}.
\end{align}
Using the same methodology as that described in the previous section and Appendix~\ref{sec:appA}, we get that $\omind = \left[\omind\right]^0 + \cct~\!\Omega_{xx} + \ct\st~\!\Omega_{xz}$, where $\left[\omind\right]^0$ is the solution in the presence of a no-slip boundary, given by Eq.~\eqref{omcleansurf} with $\vrat\to\infty$, and $\Omega_{xx}$ (resp.~$\Omega_{xz}$) is the contribution from the flow $\vec{W}_{SD}(\ex,\ex)$ (resp.~$\vec{W}_{SD}(\ex,\ez)$). The resulting rotation rate is given by
\begin{align}
	\omind = \frac{\ct\st}{8\pi\visc_1 h^3}\Bigg\{&\frac{3}{8}\left[1 +\frac{1}{2}\Eg (1 + \sst)\right] + \Eg\cct \frac{3}{\beta } \int_1^\infty{\!\!\! \int_1^\infty{\frac{F_5(2bst)}{t^3s^4}\mathrm{d}s}\mathrm{d}t}\nonumber\\
	& + \frac{1}{2\beta}F_3(2b)\left[1+\Eg(\sst-\cct)\right]\Bigg\},
\end{align}
where $b=(\vrat+1)/\beta$ and the functions $F_n$ are defined in Eqs.~\eqref{FnGnHn}-\eqref{En}. For any value of the parameters and orientation angle,  the sign of the rotation sign is that of  the prefactor $\sin(2\orient)$. As a result, a surface covered with incompressible surfactants tends to always align a swimming microorganism with the surface, similarly to the  no-slip and no-shear cases.

Furthermore, since the additional velocity field $\vec{W}$ does not have a vertical component, we see immediately that the vertical velocity induced by the interface is the same as that in the presence of a no-slip boundary, and again   a surface covered with surfactants will attract a pusher toward the interface.

In summary, the leading-order singularity modeling a swimming microorganism allowed us to generalize results known for both a solid wall and a free surface. The interface, be it a clean fluid-fluid interface, a partial slip boundary, or surface covered with surfactants always  induces  alignment parallel to the nearest surface and attraction toward it. The only common boundary conditions to these three types of interfaces is impermeability, which strongly confines the fluid in the vertical direction. This confinement provides a qualitative argument for explaining alignment and attraction of pushers near any type of boundary, as sketched in Fig.~\ref{fig:4}b. 
When tilted, a pusher will experience a higher vertical fluid force on the part of the cell closer to the interface, inducing a reorientation in the parallel direction. When swimming parallel to the interface, fluid is attracted towards the cell on its side, leading to attraction toward the surface. 
From a mathematical point of view, impermeability on the surface is enforced by a symmetric singularity as an image in the second fluid. At leading order it is therefore as if there were two symmetric micro-swimmers, which tend to align and attract each other when they are pushers.\cite{berke_2008,lauga_2009}

\subsection{Circular motion: rotlet dipole close to an interface}
\label{no}
\renewcommand{\omind}{\Omega^\text{ind}_\zz}

We turn now to a higher-order representation of a swimming bacterium in order to account for its circular motion near interfaces. Given the results above concerning the leading-order effect of the interfaces, we will assume that the the microorganism is swimming parallel to it. The micro-swimmer is now modeled by a rotlet dipole along the swimming direction, $\ex$, which can be written as a combination of  Stokes quadrupoles,
\begin{align}
	\grd(\ex,\ex)  = \frac{1}{2}\left[\gsq(\ez,\ey,\ex) - \gsq(\ey,\ez,\ex)\right].
\end{align}
As a result, the flow generated by a rotlet dipole close to a boundary can be obtained from that of the  Stokeslet, one only needs to consider the parallel Stokeslet along $\ey$ rather than $\ex$.
In this section, in order to perform dimensional analysis, we write the strength of  the rotlet dipole as  $q$ (units of N.m$^2$).

\subsubsection{Clean interface}
\newcommand{\omfunc}{\omega}

In the case of a clean fluid-fluid interface, we start by considering the problem using dimensional analysis. 
In a far-field approach, the bacterium geometry plays a role only through the non-dimensional aspect ratio $\gamma$ appearing in Fax\'en's law \eqref{eq:faxw}. The dimensional quantities involved in the rotation rate are then the rotlet dipole strength $q$, the two viscosities $\visc_1$ and $\visc_2$, and the distance $h$ to the interface. A straightforward dimensional analysis then yields a rotation rate normal to the interface given by
\begin{align}\label{DA}
	\omind = \frac{q}{\visc_1 h^4}\omfunc(\vrat,\gamma),
\end{align}
where we recall that $\lambda = \eta_2/\eta_1$. The non-dimensional function $\omfunc(\vrat,\gamma)$ needs to be determined analytically, but we can already see from Eq.~\eqref{DA} that the sign of  $\omind$, and thus the direction of rotation of the circular motion, will be determined by a comparison between  the viscosity ratio and the swimmer geometry, and will be independent of the distance to the interface $h$. Moreover, within the context of our far-field approach, we observe that this rotation rate decays as $1/h^4$. The circular motion is therefore likely to occur very close to the interface.

Starting from the image system of a Stokeslet in the presence of an interface, we derive the image system for a rotlet dipole along $\ex$, and get
\begin{align}\label{62}
	&\text{Im}\{\grd(\ex,\ex)\}=  - \grd^\star(\ex,\ex) - \frac{2\vrat}{1+\vrat}\Big[\gssd^\star(\ex,\ey,\ez) -h \gpq^\star(\ex,\ey)\Big],
\end{align}
where $\gssd$ is a stresslet dipole. Recall that the limit of a free surface (resp.~no-slip wall) is recovered in the limit of vanishing (resp.~infinite) viscosity ratio.\cite{spagnolie_2012} The structure of the image system  in Eq.~\eqref{62} reveals two roles played by   the interface: (i) a kinematic role, through the impermeability condition, that forces the existence of an opposite rotlet dipole independent of the viscosity ratio, and (ii) a viscous component, related to the balance of tangential velocity and stress. Interestingly,  in the limit of a free surface ($\vrat =0$), the viscous contribution vanishes, and the only remaining image singularity is the opposite rotlet dipole.\cite{blake_1974}

\begin{figure}[t]
	\centering
		\includegraphics{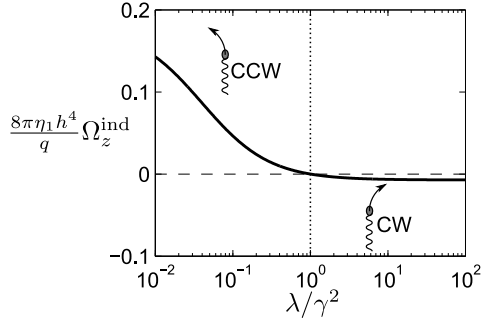}
	\caption{Interface-induced rotation rate along the normal to the interface, $\omind$, as a function of the viscosity ratio $\vrat = \visc_2/\visc_1$ divided by the aspect ratio square $\gamma^2$ (for a fixed value of $\gamma = 5$). Circular motion is predicted to be CW when $\lambda / \gamma^2 >1$ and CCW otherwise.}
	\label{fig:5}
\end{figure}

The rotation rate induced by a clean interface is then computed from Eq.~\eqref{eq:faxw} using the flow from Eq.~\eqref{62}, leading to the result
\begin{align}
	\omind = \frac{3}{256\pi}\frac{q}{\visc_1 h^4}\left(\frac{\gamma^2-1}{\gamma^2+1}+\frac{1-\vrat}{1+\vrat}\right).
	\label{eq:omegaz_1}
\end{align}
In Eq.~\eqref{eq:omegaz_1} we can identify the two different contributions which were apparent  in the image system: (i) a kinematic contribution independent of the viscosity ratio, and (ii) a viscous contribution due to the rotlet and stresslet dipoles (a potential singularity has no vorticity), independent of the micro-swimmer geometry. 
Re-writing Eq.~\eqref{eq:omegaz_1}, we get another form for the non-dimensional function $\omfunc$ in Eq.~\eqref{DA} as
\begin{align}
	\omega(\vrat,\gamma) \propto \frac{\gamma^2-\vrat}{(\vrat+1)(\gamma^2+1)}\cdot
	\label{eq:omegaz_2}
\end{align}
One can see from this expression that the sign inversion for the rotation rate, and thus the transition from CCW to CW circles, occurs for $\vrat=\gamma^2$ (see Fig.~\ref{fig:5}). As expected from the dimensional analysis, we found a condition involving the swimmer geometry and the viscosity ratio for determining the rotation direction, regardless of the distance to the interface.
For spherical swimmers ($\gamma=1$), the transition is predicted to occur exactly at a viscosity ratio of 1, but as the body becomes elongated this threshold is significantly modified. We would therefore expect a clear differentiation in rotation depending on the swimmer shape, for an identical  viscosity ratio.
%[width=0.6\textwidth]

In the presence of a clean interface between two given fluids, only the bacterium shape determines the rotation direction, and not the distance to the interface. When both fluid have comparable viscosities, the dominant behavior should be that of the free surface limit, a counter-clockwise rotation, as soon as the bacterium is elongated.
This is not consistent with experiments, where both CW and CCW rotations where observed at a free surface or a solid wall.\cite{lemelle_2010,lemelle_2013,morse_2013} This indicates a more complex role played by the interface, and motivates the next two sections.

\subsubsection{Partial slip boundary}

In the presence of a partial slip boundary, an additional length is introduced in the problem. A  dimensional analysis similar to the one carried out above yields
\begin{align}
	\omind = \frac{q}{\visc_1 h^4}\omfunc(\lgliss/h,\gamma),
\end{align}
and thus the sign of the rotation rate should depend on the distance to the wall, as opposed to the case of a clean fluid interface.

In order to analyize the effect of a partial slip boundary, we derive the solution using Blake's method directly for a rotlet dipole. Choosing $\vec{V} = -\grd^\star(\ex,\ex)$, we have in Fourier space
\begin{align}	
	\ti{\vec{w}} = \frac{\ti{\vec{w}}^0}{\ddz} + \frac{1}{4\pi\visc_1}\frac{\lgliss k_1^2\left(ik_2\ex - ik_1\ey \right)}{(\dz)(\ddz)k}e^{-k(z+h)},
\end{align}
where $\vec{w}^0=- 2\gssd^\star(\ex,\ey,\ez) +2h \gpq^\star(\ex,\ey)$.
Following an analysis similar to that detailed in Appendix~\ref{sec:appA}, we then obtain the rotation rate induced on the micro-swimmer
\begin{align}\label{68}
	\omind = \frac{3q}{64\pi\visc_1(\gamma^2+1)h^4}\Bigg\{&\frac{\gamma^2}{2} -\frac{h}{\lgliss}\left[F_4\!\!\left(\frac{h}{\lgliss}\right)+ \frac{\gamma^2-1}{2}F_5\!\!\left(\frac{h}{\lgliss}\right)\right] \nonumber\\
	&-\frac{h}{\lgliss}\left[(\gamma^2-1)G_5\!\!\left(\frac{h}{\lgliss}\right)+(3\gamma^2+1)H_5\!\!\left(\frac{h}{\lgliss}\right)\right]\Bigg\},
\end{align}
with the functions $F_n$, $G_n$, and $H_n$ defined in Eqs.~\eqref{FnGnHn}-\eqref{En}. From Eq.~\eqref{68}, the known limits of a free surface ($\lgliss\to\infty$) and solid wall ($\lgliss\to 0$) are easily recovered.

\begin{figure}[t]
	\centering
		\includegraphics{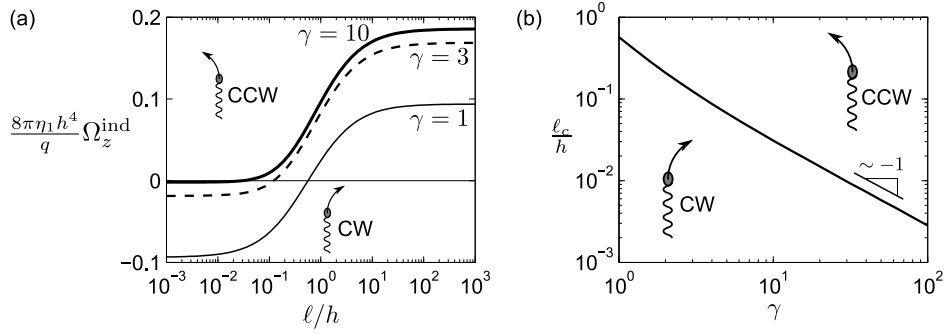}
	\caption{(a) Dependence of the perpendicular rotation rate, $\omind$, induced by a partial slip wall on a parallel rotlet dipole as a function of the normalized slip length, $\lgliss/h$, for three values of $\gamma$ (1, 3, and 10); (b) Critical value of $\lgliss/h$ at which the rotation sign changes  as a function of $\gamma$. The regions where the circular trajectories are CW and CCW are indicated on the figure.}
	\label{fig:6}
\end{figure}
%[width=0.95\textwidth]

We plot in Fig.~\ref{fig:6}(a)  the perpendicular rotation rate, $\omind$,  as a function of the normalized slip length, $\lgliss/h$, for different values of $\gamma$.  For small slip lengths compared to the distance of the cell to the wall, the effect of slip is negligible, and the standard result of a no-slip boundary is recovered (CW rotation). However, when the swimmer gets close enough to the wall, or when the slip length becomes large enough, the  rotation rate changes sign and takes that due to a free surface (CCW rotation). Since we argued above that cells are always attracted to interfaces, we thus expect  CCW  rotations to be observable in this case, consistently with the recent experimental results in the presence of slip-inducing polymers.\cite{lemelle_2013}

In  Fig.~\ref{fig:6}(b) we further  plot the dependence  of the critical dimensionless  slip length at which the rotation changes sign on the aspect ratio of the cell. We find numerically that $\lgliss/h$ scales approximatively as $1/\gamma$ at large values of the aspect ratio .

\subsubsection{Surfactant-covered interface}

In this last section we consider the case of an interface covered with incompressible surfactants. Using dimensional analysis, we obtain that the rotation perpendicular to the surface is
\begin{align}
	\omind = \frac{q}{\visc_1 h^4}\omfunc(\vrat,\beta,\gamma),
\end{align}
and therefore  the distance to the wall, $h$,  also plays a role  as it is included in $\beta$, the non-dimensional surface viscosity,

The flow generated by a rotlet dipole close to a surfactant-covered interface can be derived from the Stokeslet solution, or computed directly using Blake's method. We have  
\begin{align}
	\vec{u}_{RD}(\ex,\ex)  = \vec{u}^0_{RD}(\ex,\ex) - \frac{1}{2}(\ex.\nabla_0)(\ez.\nabla_0)\vec{W}(\ey), 
\end{align}
where $\vec{W}(\ey)$ is the additional flow field generated by a Stokeslet along $\ey$ (i.e.~Eq.~\eqref{eq:wstokeslet} with a rotation of $\pi/2$ along $\ez$). The rotation rate along the vertical axis can then be derived directly, leading to the analytical result
\begin{align}	
	\omind = \frac{3q}{64\pi\visc_1h^4}\Bigg\{&-\frac{1}{2(\gamma^2+1)}
	+\frac{1}{\beta} F_4(2b)+ \left(\Eg\right) \frac{1}{6\beta} \left[1+b(2b-1)-4b^3F_1(2b)\right]\Bigg\},
\end{align}
with the functions $F_n$ defined in Eqs.~\eqref{FnGnHn}-\eqref{En}.

\begin{figure}[t]
	\centering
		\includegraphics{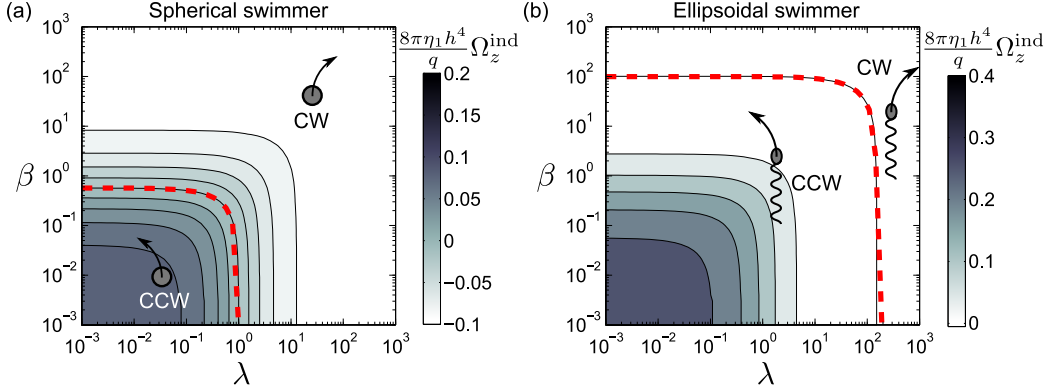}
	\caption{Dependence of the rotation rate, $\omind$, induced by an interface covered with incompressible surfactants on a parallel rotlet dipole: (a) contour values of $\omind$ for a spherical swimmer ($\gamma = 1$) and (b) for an ellipsoidal swimmer ($\gamma = 10$). The dashed line (red online) indicates the location where the  rotation rate changes sign. Regions of CW and CCW rotations are schematically shown.}
	\label{fig:7}
\end{figure}
%[width=0.95\textwidth]

The dependence of this rotation rate  with the two non-dimensional viscosities, $\lambda = \visc_2/\visc_1$ and $\beta = \visc_s/h\visc_1$, is shown in Fig.~\ref{fig:7}. 
For low cell aspect ratio, the  sign of the circular rotation is similar to the no-slip case for a wide range of parameters  (CW rotation), and thus opposite to the prediction in the case of clean interface. 
The sign of rotation is then changing  at low $\beta$ (low surfactant concentration) and low viscosity ratio $\lambda$. Furthermore, as the aspect ratio of the cell $\gamma$ increases, the region displaying  CCW rotation becomes larger, allowing for both rotations to   be potentially observed experimentally in cell populations of different sizes or on surfaces with fluctuations in surfactant concentration.

Comparing with other interfacial models, we get that when $\beta$ tends to infinity, the solid wall limit is recovered. Furthermore, and despite the fact that the surfactant model is not supposed to recover   exactly the clean-interface limit, we see that the clean-interface threshold for rotation inversion ($\lambda = \gamma^2$), corresponds to the order of magnitude of the vertical asymptote of the dashed line in 
Fig.~\ref{fig:7}.

\begin{figure}[t]
	\centering
		\includegraphics{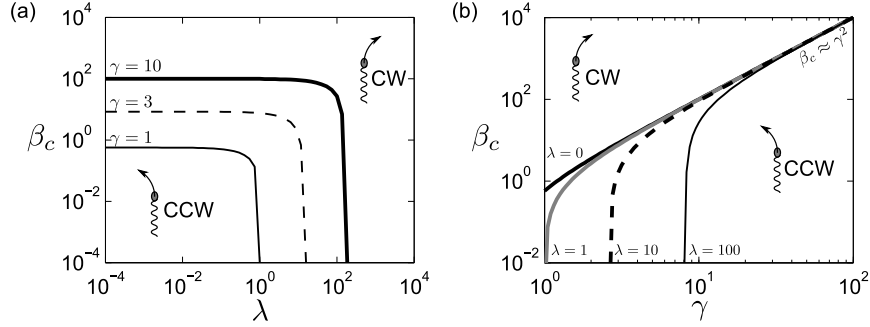}
	\caption{Critical value of the dimensionless surface viscosity, $\behat$, at which the rotation changes sign, as a function of $\lambda$ for three values of $\gamma$ (a), and as a function of $\gamma$ for four values of $\lambda$ (b). Regions of CW and CCW rotations are indicated on the figure.}
	\label{fig:8}
\end{figure}
%[width=0.95\textwidth]

We then plot in Fig.~\ref{fig:8}  the critical value of $\beta$ at which there is a change in sign of the rotation rate, as a function of the viscosity ratio $\lambda$ for different aspect ratios (Fig.~\ref{fig:8}a), and as a function of $\gamma$ for different $\lambda$ (Fig.~\ref{fig:8}b). For large cell aspect ratios, we find that $\beta$ scales as $\gamma^2$, similarly to the scaling seen for  $\lambda$. This can be interpreted by noting that this problem is the same as that in the presence of a clean interface, but for the presence of a third fluid, and thus a third viscosity that needs to be compared with the viscosity where the swimming microorganism is located. Hence, the criterion for a change of sign in the rotation direction is similar   for both non-dimensional viscosities, $\vrat$ and $\beta$.

{There is in general a wide range of surface viscosities, depending on multiple parameters, such as the type of surfactants or temperature.\cite{fuller_2012} For instance, for a bacterium swimming at a typical distance of 1~$\mu$m from the interface,  $\beta$ can be of order  $10^6$ when the interface is a monolayer at a water-air interface.\cite{brooks_1999}}
As a result, CW rotation (as in the no-slip case) is more likely to be observed. In contrast, in the case of   an amphiphilic bilayer,\cite{shkulipa_2005} {the values can be much smaller, $\beta =O(1)$.} As a result, there is likely  a wide range of parameters  where both CW and CCW rotation could be observed in a population of cells.  Recent experimental results on contaminated free surfaces showed that both CW and CCW circular motion could be seen.\cite{lemelle_2010,morse_2013} By considering the presence of surfactants on the surface, our analysis  shows that the high viscosity fluid film at the interface could indeed alter the natural shear-free rotation direction and lead to CW motion.

\section{Conclusion}
\label{sec:ccl}

In this paper we have used a far-field hydrodynamic approach to model the surface swimming of bacteria employing helical flagella. The motivation for this work was the discrepancy between theoretical predictions and experimental observations. Specifically,
theory predicts that near a rigid wall the cells should always display CW motion, whereas recent experiments where polymers were used to induce slip at the wall showed that rotation in the opposite direction was possible. Similarly, cells should rotate in a CCW motion at a free surface whereas if surfactants are present experiments show that CW motion is also observed.

To develop a model we have represented the helical swimmer as a superposition of hydrodynamic singularities and investigated its hydrodynamic interactions with three types of surfaces: a clean fluid-fluid interface, a rigid wall with a finite slip length, and an interface covered by incompressible surfactants.  The leading-order singularity in the flow field of the cell  is a Stokes dipole (stresslet), characterized by a $1/r^2$ spatial decay. The interactions between that singularity and all three types of surfaces systematically lead to a reorientation of the swimming cells parallel to,  and an attraction by,  the surface. Circular motion of the cells are due to wall effects on  a higher-order singularity, namely a rotlet dipole, which decays spatially as $1/r^3$.  In that case, the specific boundary conditions at the interface, together with the shape of the cell, play a crucial role in determining the direction of rotation of the cell, and transitions between CW and CCW are predicted to take place in similar experimental setups. 

Our results indicate thus that the recent experimental finding on transition in rotation direction can be understood  as the consequence of complex boundary conditions on the nature of hydrodynamic interactions between the swimming cells and the surfaces. The main assumption made in our paper is that we have only considered the leading-order hydrodynamics effects for all influences of the interfaces (attraction and rotation). 
This is, admittedly, a severe assumption which is expected to break down as soon as the cell is within about one body length from the interface. In order to obtain more quantitative predictions, one would then need to either include the effect of higher-order singularities, or resort to a fully computational approach. The advantage of our approach however is that it allows us to identify the fundamental physical process at play in setting the direction of rotation, and that it is expected to remain valid generically for all cells exploiting helical swimming. Our findings could potentially be exploited in a numbers of ways, for example surface swimming could be used as a proxy for determining the rheological properties of the nearby interface or to selectively stir or sort individual cells from bacterial populations. We hope that our study will motivate further work along these directions.

\section*{Acknowledgements}
The authors thank the Department of Mechanical and Aerospace Engineering at the University of California, San Diego where this work was initiated. 

\appendix

\renewcommand{\omind}{\Omega^\text{ind}}

\section{Known Fourier transforms}
\label{sec:appB}

We reference here some useful Fourier transforms
\begin{align}
	\four{\frac{1}{r}} = \frac{1}{k}e^{-kz},\quad \four{\frac{1}{r^3}} = \frac{1}{z}e^{-kz},\quad \four{\frac{x}{r^3}} = \frac{ik_1}{k}e^{-kz}
\end{align}
where $r^2 = x^2+x^2+z^2$.

\section{Reorientation of a Stokes dipole close to a partial slip boundary}
\label{sec:appA}

We present here the details of the derivation for  the rotation rate, $\omind_\yy$, induced by a partial slip boundary on a tilted Stokes dipole. The method is  general, and the same procedure is applied throughout the paper for deriving the rotation rate induced by a complex interface. In Fourier space, the rotation rate induced by a flow $\ti{\vec{u}}$ on a prolate spheroid of aspect ratio $\gamma$, oriented along $\ee$, reads
\begin{align}
	\ti{\vec{\Omega}}\left[\ti{\vec{u}}\right] &= \frac{1}{2}
	\begin{pmatrix}
	-ik_2\ti{u}_z-\derp{\ti{u}_2}{z}\\
	\derp{\ti{u}_1}{z}+ik_1\ti{u}_z\\
	ik_2\ti{u}_1-ik_1\ti{u}_2
	\end{pmatrix} \nonumber\\
	 &\quad+ \frac{1}{2}\Eg
	\begin{pmatrix}
	\ct\st[ik_1\ti{u}_2+ik_2\ti{u}_1]-\sst\left[\derp{\ti{u}_2}{z}-ik_2\ti{u}_z\right]\\
	[\sst-\cct][\derp{\ti{u}_1}{z}-ik_1\ti{u}_z]-2\ct\st\left[ik_1\ti{u}_1+\derp{\ti{u}_z}{z}\right]\\
	\ct\st[\derp{\ti{u}_2}{z}-ik_2\ti{u}_z]-\cct\left[ik_1\ti{u}_2+ik_2\ti{u}_1\right]
	\end{pmatrix}.
	\label{eq:om_gen}
\end{align}
The tilted Stokeslet dipole has three contribution, a stresslet, $\vec{u}_{SS}(\ex,\ez)$, and two Stokeslet dipoles, $\vec{u}_{SD}(\ex,\ex)$ and $\vec{u}_{SD}(\ez,\ez)$.

\subsection{Stresslet contribution}
\label{sec:StressletContribution}
We compute first the stresslet contribution. For a stresslet $\vec{u}_{SS}(\ex,\ez)$, we have $\ti{\vec{w}} = \ti{\vec{w}}^1 + \ti{\vec{w}}^2$, where
\begin{align}
	\ti{\vec{w}}^1 = \frac{\ti{\vec{w}}^0}{\ddz}, \quad \ti{\vec{w}}^2 = \frac{\lgliss}{4\pi\visc_1(\dz)(\ddz)}\left(\frac{k_2^2}{k}\ex - \frac{k_1k_2}{k}\ey\right)e^{-k(z+h)}.
\end{align}
The coefficients of the no-slip solution read
\begin{align}
	&A_1^0 = 2\frac{k-k_1^2h}{k}e^{-kh},\quad A_2^0 = 2h\frac{-k_1k_2}{k}e^{-kh},\quad A_z^0 = 0,\quad B^0 = \frac{2ik_1}{k}(1-kh)e^{-kh}.
\end{align}
We note $\ti{\Omega}^1=\ti{\Omega}_\yy\left[\ti{\vec{w}}^1\right]$ and $\ti{\Omega}^2=\ti{\Omega}_\yy\left[\ti{\vec{w}}^2\right]$.
From the expression of the rotation rate in Eq.~\eqref{eq:om_gen}, we have $\ti{\Omega}^1 = {\ti{\Omega}^0}/{(\ddz)}$.
From Eq.~\eqref{eq:ufour}, we note that $\ti{\vec{w}} = \ti{\vec{w}}_a + z \ti{\vec{w}}_b$, and thus $\ti{\Omega} = \ti{\Omega}_a + z\ti{\Omega}_b$. This decomposition holds in real space,\cite{lauga_2005} leading to
\begin{align}
	\left(1-2\lgliss\derp{}{z}\right)\Omega_a^1 &= \Omega^0_a, \quad \left(1-2\lgliss\derp{}{z}\right)\Omega_b^1 = \Omega^0_b.
\end{align}
The resulting differential equation reads
\begin{align}
	\left(1-2\lgliss\derp{}{z}\right)\Omega^1& = \Omega^0 - 2\lgliss \Omega_b^1.
\end{align}
We decompose then $\Omega^1$ in two terms, $\Omega^1 = \Omega^{01} + \Omega^{11}$, so that
\begin{align}
	\left(1-2\lgliss\derp{}{z}\right)\Omega^{01} &= \Omega^0,\quad \left(1-2\lgliss\derp{}{z}\right)\Omega^{11} = -2\lgliss \Omega^1_b.
\end{align}
This last equation can be rewritten as
\begin{align}
	\left(1-2\lgliss\derp{}{z}\right)^2\Omega^{11}  = -2\lgliss\Omega_b^0.
\end{align}
The first term, $\Omega^{01}$, can be computed directly knowing the no-slip solution. However, the second term is not straightforward, and needs to be determined in Fourier space, as well as $\Omega^2$.  
We have then
\begin{align}
	\ti{\Omega}_b^0 &= \Eg \frac{B^0}{8\pi\visc_1}\left[(\cct-\sst)ik_1k+ \st\ct(k_1^2+k^2)\right] e^{-kz}, \\
	\ti{\Omega}^2 &=-\frac{1}{8\pi\visc_1}\frac{\lgliss k_2^2 e^{-k(z+h)}}{(\dz)(\ddz)}\left[1+\Eg\left(\sst-\cct + 2\ct\st\frac{ik_1}{k}\right)\right],
	\label{eq:bug}
\end{align}
These expressions can be inverted, as all contributions are known Fourier transforms (see Appendix \ref{sec:appB}). We have therefore $\Omega = \Omega^{01} + \Omega^{11} + \Omega^2$, with
\begin{align}
	&\left(1-2\lgliss\derp{}{z}\right)\Omega^{01} = \Omega^0, \\
	&\left(1-2\lgliss\derp{}{z}\right)^2\Omega^{11}  = -\frac{\lgliss}{4\pi\visc_1} \Eg \left(1+h\derp{}{z}\right)\bigg[ 2(\sst-\cct)\derp{}{z}\left(\frac{3x^2}{\r^5}-\frac{1}{\r^3}\right) \nonumber \\
	& \qquad \qquad \qquad \qquad \qquad \qquad \qquad \qquad \qquad \qquad \quad + \sin(2\orient)\left(\derp{^2}{z^2} - \derp{^2}{x^2}\right)\left(\frac{x}{\r^3}\right) \bigg],\\
	&\left(1-\lgliss\derp{}{z}\right)\left(1-2\lgliss\derp{}{z}\right)\Omega^2 = \frac{\lgliss}{8\pi\visc_1} \left[1+\Eg(\sst-\cct)\right]\derp{}{z}\left(\frac{1}{\r^3}-\frac{3y^2}{\r^5}\right)\nonumber\\
	&\qquad \qquad \qquad \qquad \qquad \qquad\quad +\frac{\lgliss}{8\pi\visc_1} \Eg\sin(2\orient)\derp{^2}{y^2}\left(\frac{x}{\r^3}\right),
\end{align}
where $\r^2 = x^2+y^2+\z$,  with $\z = z+h$. Since we are looking for the solution at $x=y=0$, we need to integrate the following equations
\begin{align}
	&\left(1-2\lgliss\derd{}{\z}\right)\Omega_{SS}^{01} =  \frac{3}{8\pi\visc_1} \frac{1}{\z^3}\left[\frac{2h}{\z}-1 + \Eg(\sst-\cct)\left(1-\frac{8h}{\z} +\frac{8h^2}{\z^2}  \right)\right], \\
	&\left(1-2\lgliss\derd{}{\z}\right)^2\Omega_{SS}^{11}  = -\frac{3\lgliss(\sst-\cct)}{2\pi\visc_1} \Eg  \frac{1}{\z^4}\left(1-\frac{4h}{\z}\right),\\
	&\left(1-\lgliss\derd{}{\z}\right)\left(1-2\lgliss\derd{}{\z}\right)\Omega_{SS}^2 = -\frac{3\lgliss}{8\pi\visc_1} \left[1+\Eg(\sst-\cct)\right] \frac{1}{\z^4}\cdot
\end{align}
The final rotation rate due to $\vec{w}$ reads $\Omega_{SS}^w = \Omega_{SS}^{01} + \Omega_{SS}^{11} + \Omega_{SS}^2$.

\subsection{Parallel Stokes dipole contribution}
\label{sec:ParallelStokesDipoleContribution}

For the parallel Stokes dipole, $\vec{u}_{SD}(\ex,\ex)$, we have 
\begin{align}
	&\vec{V}=\gsd^\star(\ex,\ex),\quad \vec{w}^0 = -2\gsd^\star(\ex,\ex) + 2h\gsq^\star(\ez,\ex,\ex) -2h^2\gpq^\star(\ex,\ex),\\
	&\ti{\vec{w}} = \ti{\vec{w}}^1+\ti{\vec{w}}^2,\quad \ti{\vec{w}}^1 = \frac{\ti{\vec{w}}^0}{\ddz}, \quad \ti{\vec{w}}^2 =  -\frac{\lgliss ik_1k_2 e^{-k(z+h)}}{2\pi\visc_1(\dz)(\ddz)k^2}\left[k_2\ex-k_1\ey\right].
\end{align}
The corresponding coefficients in Fourier space are given by
\begin{align}
	&A_1^0 = -\frac{2ik_1}{k^3}\left(k_1^2(1-kh) + 2k_2^2\right)e^{-kh},\quad A_2^0 = \frac{2ik_1^2k_2}{k^3}(1+kh)e^{-kh},\quad A_3^0 = 0,\\
	&B^0 = \frac{2k_1^2}{k^2}(1-kh)e^{-kh}.
	\label{eq:sdcoefpar}
\end{align}

Following the same procedure, we find that the rotation rate, $\omind_\yy$, induced by a parallel Stokes dipole $\vec{u}_{SD}(\ex,\ex)$ is given by $\Omega_{SDX} = \Omega_{SDX}^V+ \Omega_{SDX}^{01} + \Omega_{SDX}^{11} + \Omega_{SDX}^2$ where
\begin{align}
	&\Omega_{SDX}^V = -\frac{3\ct\st}{8\pi\visc_1}\Eg \frac{1}{\z^3},\\
	&\left(1-2\lgliss\derd{}{\z}\right)\Omega_{SDX}^{01} = \frac{3\ct\st}{4\pi\visc_1}\Eg \frac{1}{\z^3}\left(1-\frac{7h}{\z} +\frac{7h^2}{\z^2}\right), \\
	&\left(1-2\lgliss\derd{}{z}\right)^2\Omega_{SDX}^{11}  = -\frac{21\lgliss\ct\st}{8\pi\visc_1}\Eg \frac{1}{\z^4}\left(1-\frac{4h}{\z}\right),\\
	&\left(1-\lgliss\derd{}{\z}\right)\left(1-2\lgliss\derd{}{\z}\right)\derd{\Omega_{SDX}^2}{\z} = -\frac{3\lgliss\ct\st}{2\pi\visc_1}\Eg \frac{1}{\z^5}\cdot
\end{align}

\subsection{Perpendicular Stokes dipole contribution}
\label{sec:PerpStokesDipoleContribution}
For the perpendicular Stokes dipole, $\vec{u}_{SD}(\ez,\ez)$, we have
\begin{align}
	&\vec{V}=\gsd^\star(\ez,\ez),\quad \ti{\vec{w}} = \frac{\ti{\vec{w}}^0}{\ddz}, \quad B^0 = 2(kh-1)e^{-kh},\\
	&\vec{w}^0 = 2\left[-\gsd^\star(\ez,\ez) + h\gsq^\star(\ez,\ez,\ez) +2h\gpd^\star(\ez)-h^2\gpq^\star(\ez,\ez)\right].
\end{align}
The rotation rate, $\omind_\yy$, induced by a perpendicular Stokes dipole $\vec{u}_{SD}(\ez,\ez)$ is given by $\Omega_{SDZ} = \Omega_{SDZ}^V+ \Omega_{SDZ}^{01} + \Omega_{SDZ}^{11}$, with
\begin{align}
	\Omega_{SDZ}^V &= \frac{3\ct\st}{4\pi\visc_1}\Eg\frac{1}{\z^3},
	\\
	\left(1-2\lgliss\derd{}{\z}\right)\Omega_{SDZ}^{01} &= -\frac{3\ct\st}{2\pi\visc_1}\Eg \frac{1}{\z^3}\left(1-\frac{6h}{\z} +\frac{6h^2}{\z^2}\right), \\
	\left(1-2\lgliss\derd{}{\z}\right)^2\Omega_{SDZ}^{11} & = \frac{9\lgliss\ct\st}{2\pi\visc_1}\Eg \frac{1}{\z^4}\left(1-\frac{4h}{\z}\right)\cdot
\end{align}

\section{Typical differential equations and solutions}
\label{sec:TypicalDifferentialEquationsAndSolutions}

The differential equations in $\z$ giving the rotation rate and velocities induced by the nearby boundary are of the three following types
\begin{align}
	&\left(1-2\lgliss\derd{}{\z}\right)f_1 =  \frac{1}{\z^n}, \\
	&\left(1-2\lgliss\derd{}{\z}\right)^2f_2  = \frac{1}{\z^n},\\
	&\left(1-\lgliss\derd{}{\z}\right)\left(1-2\lgliss\derd{}{\z}\right)f_3 = \frac{1}{\z^n},
\end{align}
where $n$ is a positive integer. We keep here the coefficients corresponding to the partial slip case, however similar equations are obtained in the case of a surfactant-covered interface. Knowing that the solution must vanish at infinity, the solutions for these equations read
\begin{align}
	f_1(\z) &= \frac{1}{2\lgliss}\frac{e^{\z/2\lgliss}}{\z^{n-1}}E_n\!(\z/2\lgliss),\\
	f_2(\z) &= \frac{1}{4\lgliss^2}\frac{e^{\z/2\lgliss}}{\z^{n-2}} \int_1^\infty{\frac{E_n\!(\z t /2\lgliss)}{t^{n-1}}\mathrm{d}t},\\
	f_3(\z) &= \frac{1}{2\lgliss^2}\frac{e^{\z/\lgliss}}{\z^{n-2}} \int_1^\infty{\frac{e^{-\z t/2\lgliss}}{t^{n-1}}E_n\!(\z t /2\lgliss)\mathrm{d}t},
\end{align}
where $E_n$ is the exponential integral function of order $n$ defined in Eq.~\eqref{En}.

\end{document}